 \documentclass[preprint2,aps,floats,epsfig]{aastex}

\newcommand{\beq}{\begin{equation}}
\newcommand{\eeq}{\end{equation}}
\newcommand{\beqa}{\begin{eqnarray}}
\newcommand{\eeqa}{\end{eqnarray}}
\newcommand{\mx}{\left[\begin{array}}
\newcommand{\finmx}{\end{array}\right]}
\newcommand{\mxp}{\left(\begin{array}}
\newcommand{\finmxp}{\end{array}\right)}
\newcommand{\casos}{\left\{\begin{array}}
\newcommand{\fincasos}{\end{array}\right.}
\newcommand{\rcasos}{\left.\begin{array}}
\newcommand{\rfincasos}{\end{array}\right\}}
\newcommand{\lsim}{\lesssim}
\newcommand{\gsim}{\gtrsim}

\newcommand{\halff}{\frac{1}{2}}

\newcommand{\rres}{R_{\rm res}}

\shorttitle{Pulsar acceleration by sterile neutrinos emission}
\shortauthors{Nardi and Zuluaga}

\begin{document}


\title{
Pulsar acceleration by asymmetric emission 
of sterile neutrinos  } 

\author{Enrico Nardi\altaffilmark{1}}
\affil{Theory Division, CERN
CH-1211 Geneva 23, Switzerland}
\email{enardi@naima.udea.edu.co}
\and
\author{Jorge I. Zuluaga}
\affil{Departamento de F\'\i sica,
 Universidad de Antioquia,  
A.A. 1226, Medell\'\i n,
 Colombia}
\email{jzuluaga@naima.udea.edu.co}

\altaffiltext{1}{Permanent address: Departamento de F\'\i sica,
 Universidad de Antioquia, A.A. 1226, Medell\'\i n, Colombia}

\begin{abstract}
  A convincing explanation for the observed pulsar large peculiar
  velocities is still missing. We argue that any viable particle
  physics solution would most likely involve the resonant production
  of a non-interacting neutrino $\nu_s$ of mass $m_{\nu_s}\sim
  20$--$50\>$ keV. We propose a model where anisotropic magnetic field
  configurations strongly bias the resonant spin flavour precession of
  tau antineutrinos into $\nu_s$.  For internal magnetic fields
  $B_{\rm int} \gsim 10^{15}\,$G a $\bar\nu_\tau$-$\nu_s$ transition
  magnetic moment of the order of $10^{-12}$ Bohr magnetons is
  required.  The asymmetric emission of $\nu_s$ from the core can
  produce sizeable natal kicks and account for recoil velocities of
  several hundred kilometers per second.
\end{abstract}



\section{INTRODUCTION} 

Measurements of the pulsars proper motion 
have shown that  the pulsar population
is characterized by peculiar velocities 
much higher than those of all other 
stellar populations \citep{gunn1970,lyn1994,han1997,cor1997,lor1997f}.  
While the  precise form of the velocity distribution 
is uncertain, recent studies indicate that 
it must extend to  $\sim 1000\,$km s$^{-1}\,$,  
with a mean of $\sim 500$ km s\,$^{-1}$ \citep{lor1997}. 
Since the velocities of pulsars' likely progenitors, 
the massive stars, are among the lowest in the galaxy
($\sim 10$--$40$ km s$^{-1}$) it is  natural to think that the mechanism 
responsible for their acceleration can be 
related either to supernova (SN) core collapse and explosion,  
or to the post-explosion cooling phase when, within a few 
seconds, a huge amount of energy is emitted from the 
proto-neutron star (PNS) through the radiation of  $\sim 10^{58}$ 
neutrinos of all flavours.  

To date, no convincing physical process that could account for pulsar
acceleration has been identified (a review of different kick
mechanisms has been recently given by Lai (1999)).  The astrophysical
solutions that have been proposed include, in broad terms, the
possibility of symmetric and of asymmetric SN explosions. In the first
case, large recoil velocities are achieved in response to the mass
loss in the (symmetric) explosion of one component of a close binary
system \citep{got1970f,ibe1996-1997}.  It was shown that the remnant
can achieve velocities comparable to the orbital velocity of the SN
progenitor.  However, detailed Monte Carlo simulations of binaries'
evolution through the SN phase fail to reproduce the observed velocity
distributions of pulsars and hint at asymmetric explosions
\citep{lor1997,dew1987,heu1997,hug1999}.  In the second case,
asymmetrical stellar collapse and explosion occurs because of
pre-explosion instabilities and core density distortions, imparting to
the residue an intrinsic kick \citep{jan1994,bur1995,bur1996}.  The
momentum scale of the mass ejecta in the SN explosion is of the order
of $\approx 6\times 10^{53}$ erg/$c$, corresponding to about ten solar
masses ejected with velocities $\approx 10^4$ km s$^{-1}$.  Then, to
accelerate the remnant of about $1.5\, M_\odot$ to the largest
observed velocities, it is necessary to generate a momentum asymmetry
in the ejecta of the order of a few per cent.  The asymmetric
explosion scenario remains unsatisfactory in the sense that initial
anisotropies in the collapsing core sufficiently large to generate
asymmetries of this size have to be input artificially \citep{bur1996}
and, from the point of view of present (2-D) hydrodynamic
calculations, are difficult to justify \citep{jan1994}.

More recently it has been suggested that a 
completely different class of mechanisms could 
be responsible for imparting a natal kick to 
the neutron star. Virtually all the gravitational 
binding energy released during the collapse of the 
progenitor iron core ($\approx 3\times 10^{53}$ ergs)
is emitted during the first 10 seconds after core bounce     
in the form of neutrinos and antineutrinos. 
The momentum in neutrinos is therefore comparable to the 
momentum in the ejected matter. Hence, also in this case 
a few per cent asymmetry in the neutrino emission  could 
suffice to account for the largest velocities observed.
The particle physics explanations for 
the pulsar acceleration mechanisms 
can be classified according to the 
underlying mechanism at the origin of the 
momentum asymmetry in the  neutrino emission.
The asymmetry could be in the {\it number} of neutrinos, 
or in the neutrino {\it energy}
(since the number of neutrinos is not conserved when neutrinos stream
out of the core, this only refers to the microscopic cause of the
asymmetry, and does not describe the macroscopic effect).  In both
cases, the presence of a strong magnetic field $B\sim
10^{15}$--$10^{16}\,$G provides the asymmetric initial condition for
the system.  In the first case, parity violation in electroweak
interactions was thought to be at the origin of asymmetric neutrino
emission from strongly magnetized PNS.  The cumulative effect of
multiple neutrino scatterings off the slightly polarized nucleons
would result in a sizeable anisotropy in the neutrino momentum
\citep{hor-pi1998,hor-li1998,lai1998a}.  More detailed studies showed
that this effect was largely overestimated \citep{kus1998f,arr1999}.
When neutrinos are in local thermodynamic equilibrium (as they are to
a very good approximation in the interior of a PNS) detailed balance
ensures that no cumulative effect from multiple scatterings can build
up.  
Local effects in the PNS atmosphere, as for example the 
deviations from thermal equilibrium that occurr close
to the neutrinosphere, can generate asymmetries
in the neutrino emission. However, to date detailed analyses 
of these effects have failed to produce kicks as large as 
1000 km s$^{-1}$ \citep{arr1999}.
The second case relies on the assumption
that resonant transitions (oscillations \citep{kus1996-1997,gra1998f}
or spin-flavour precession \citep{akh1997f}) between different
neutrino flavours occur inside the PNS.  The resonance surface for the
conversion becomes the effective surface of last scattering (the
neutrinosphere) for the neutrino flavour with the largest mean free
path.  Since a strong magnetic field can slightly distort the
resonance surface, neutrinos are emitted from regions with different
temperatures, which yields an asymmetry in the overall momentum.  Also
in this case, the real effect was largely overestimated.  This is
because the variation of the temperature over the effective
neutrinosphere cannot be directly related to an anisotropy in the
neutrino energy.  The energy flow from the star is governed by the
inner core emission of neutrinos; in first approximation, local
processes in the PNS atmosphere, where the resonant conversion takes
place, cannot modify the (isotropic) neutrino emission from the core
\citep{jan1999}.

In conclusion, both the macroscopic parity violation and the resonant
neutrino conversion scenario, which for a while seemed to provide
elegant particle physics mechanisms to account for pulsars'
acceleration, did not survive more detailed analysis. However, the
study of these attempts and of the reasons why they fail to produce
natal kicks of the right magnitude can guide us to infer some of the
conditions that must be satisfied by a viable particle physics
solution.

\begin{itemize}

\item
The neutrino observations from SN1987A, and in particular the duration
of the $\bar\nu_e$ burst, support the theoretical expectation that
active neutrinos are copiously produced inside the PNS core and that
they are the main agents of the energy emission.  Theoretical studies
indicate that, to a good approximation, the total energy carried away
by neutrinos is equipartitioned among the six neutrino and
antineutrino flavours \citep{jan1995}.  This implies that conversions
between the active flavours would simply result in exchanging the
respective spectra, and cannot affect the main characteristics of
energy emission \citep{jan1999}.  We learn from this that if neutrino
conversion is responsible for the kicks, then the conversion must be
into some new particle that does not interact with matter in the same
way than the standard neutrinos: most probably a very weakly
interacting or sterile neutrino $\nu_s$.

\item 
In thermal equilibrium, the microscopic peculiarities of neutrino
scattering processes cannot build up any macroscopic asymmetry. This
applies also to possible neutrino scattering into new exotic states.
This hints at some macroscopic condition anisotropically realized
inside the PNS.  If this has to influence the neutrino conversion
rates, we are naturally led to thinking at some asymmetrically located
resonance, or at local violation of the adiabaticity conditions for
the conversion.

\item
An anisotropic resonant conversion of an active neutrino into an
exotic state is still not sufficient to generate a momentum asymmetry.
Close to the resonance region the parent $\nu$ and the $\nu_s$ would
both have an asymmetric momentum distribution, but their asymmetries
would compensate in the total momentum. However, a non-interacting
$\nu_s$ would freely escape from the PNS, thus preserving its original
asymmetry, while the active neutrino keeps interacting with the
background particles.  If $\nu$ has enough time to recover a symmetric
momentum distribution before escaping from the star, then the $\nu_s$
asymmetry will be left unbalanced.  This suggests that the conversion
should occur far inside the $\nu$ neutrinosphere, and most likely in
the high-density regions close to the core.

\end{itemize}

The picture emerging from these considerations is a resonant
conversion of an active flavour into a non-interacting neutrino, at
the large densities and temperatures typical of the PNS regions close
to the core.  In turn, matching the resonance condition hints at a
sterile neutrino mass $\approx 20$--50 keV.  In order to implement the
previous scheme it is important to identify all the possible
macroscopic asymmetries of the system.  As we will discuss in the next
section, there are observational indications that, in some cases,
pulsar magnetic fields are better described by an off-centred dipole,
rather than simply by an oblique dipole with the centre of the
magnetic axis coinciding with the centre of the star.  Clearly this
implies anisotropic conditions inside the PNS core, since the magnetic
field strength can easily vary by more than one order of magnitude
within regions of equal matter density.  Then it is not difficult to
envisage how in such an anisotropic background the
Mikheyev--Smirnov--Wolfenstein (MSW) mechanism for matter-enhanced
neutrino oscillations \citep{wolf1978,mik1985} or resonant
spin-flavour precession (RSFP) \citep{akh1988a-1988b,lim1988} could
result in large asymmetries in the rate of neutrino conversion between
opposite hemispheres. Moreover, if the conversion involves
non-interacting neutrinos, they will freely escape from the core,
yielding a large momentum asymmetry.  We stress that the solution we
are proposing here is intrinsically different from previously studied
mechanisms that exploit resonant neutrino conversion and emission from
a slightly deformed resonance surface
\citep{kus1996-1997,gra1998,akh1997}.  In our case, it is the
conversion probability that is not isotropically distributed inside
the (spherical) region where the resonant condition is satisfied.

In Section 2 we will review some of the arguments in support of our
basic assumption that the magnetic field in the PNS interior can be
characterized by sizeable anisotropies. In Section 3 we will briefly
review the physics of RSFP.  For definiteness, we will assume that tau
antineutrinos can be resonantly converted into sterile neutrinos
$\nu_s$.  We will show that for acceptable values of the PNS magnetic
field and of the neutrino transition magnetic moment $\mu_\nu$ the
adiabaticity conditions for an efficient conversion are satisfied in
regions asymmetrically displaced from the star centre.  A MSW solution
to the problem can also be implemented straightforwardly, since the
asymmetrically polarized background results in an anisotropic neutrino
potential \citep{nun1997}.  However, this requires more fine tuning in
the choice of the relevant parameters and we will only comment briefly
on this possibility.  The details of the PNS model we have used and of
neutrino diffusion from the core will be described in Section 4.
Finally in Section 5 we will present our results and draw the
conclusions.

\section{EVIDENCE FOR ANISOTROPIC MAGNETIC FIELDS} 

The evidence in support of asymmetric magnetic field topologies for
variable magnetic stars and pulsars was discussed long ago by Harrison
and Tademaru \citep{har1975,tad1975,tad1976}.  They argued that
pulsars' magnetic fields can be well modeled by an off-centred
magnetic dipole, displaced from the centre of the star and oriented
obliquely with respect to the rotation axis in an arbitrary direction.
They also showed that such a field configuration radiates
asymmetrically low-frequency electromagnetic radiation, and they tried
to explain through this mechanism `post-natal' pulsar acceleration at
the expense of the star rotational energy.  However, velocities much
in excess of 100 km s$^{-1}$ cannot be explained in this way, since
for typical rotational periods of the order of $10$ ms the rotational
energy does not exceed by much the total kinetic energy of the star.
Moreover, the lack of any observed correlation between the magnetic
field and the transverse velocity, or between the direction of the
motion and the magnetic axis \citep{lor1997,des1999f} has virtually
ruled out the post-natal acceleration model.  Asymmetric magnetic
field configurations can also affect $\nu_e$ and $\bar\nu_e$
absorption cross sections on neutrons and protons. Lai and Qian
(1998b) studied the possible asymmetry in the neutrino emission that
can be generated in this way, and showed that the effect is again too
small to produce appreciable kicks.

However, an anisotropic magnetic field can still be responsible for
pulsar acceleration by producing sizeable asymmetries in the neutrino
emission during the early PNS cooling phase.  In principle, the
mechanism we want to propose can be effective independent of the
particular type of magnetic field model, as long as it predicts
sizeable anisotropies in the field strength. However, for definiteness
we will concentrate on the simple off-centred and decentred magnetic
dipole models \citep{har1975,tad1975,tad1976} also because they are
supported by a few observational evidences.  Decentred dipole models
were first introduced in the attempt to improve the representation of
the fields of periodic magnetic variable stars \citep{lan1970}.  It
was later suggested that these models could also account for special
features in the electromagnetic emission of the so-called interpulse
pulsars, which are characterized by secondary interpulses well
separated from the main pulse, and of an intensity up to two orders of
magnitude smaller \citep{han1986}.  A displacement of the magnetic
dipole along the magnetic axis of an amount $\delta$ in units of
stellar radius would produce unequal surface field intensities in the
ratio $(1+\delta)^3/(1-\delta)^3$, providing a simple explanation for
the relative weakness of the secondary interpulse.  The decentre
parameter $\delta$ inferred from observational studies of variable
magnetic stars and interpulse pulsars ranges between 0.1 and 0.6
\citep{har1975,tad1975,tad1976}.  In some cases surface fields of
opposite polarities, separated by much less than 180$^{\rm o}$ have
been observed \citep{han1986}.  Asymmetries of this kind are not
explained by the simple decentred model, and suggest that the dipole
is actually {\it off-centred}. Namely, the magnetic dipole $\mu=
(\mu_z,\mu_\rho,\mu_\phi)$ (in cylindrical coordinates) is arbitrarily
oriented and displaced from the centre of the star by an amount
$\delta_z$ along the spin axis and $\delta_\rho$ in the radial
direction. Since the dipole axis does not intersect the spin axis, the
angular separation between the main and the secondary pulses is easily
accounted for.

The study of interpulse pulsars is statistically limited by the small
sample of a few per cent of all pulsar population.  This is likely to
be due to the fact that both the angle of inclination between the
rotation axis and the line of sight, and the angle between the spin
and the magnetic axis, must be close to 90$^{\rm o}$ to allow for
their detection.  Still, the few direct observations available today
suggest that it is not unreasonable to assume that most pulsars have
oblique and off-centred dipole fields, or possibly some more
complicated field configuration implying sizeable anisotropies in the
magnetic field topology.

Since off-centred dipoles radiate energy at a different rate with
respect to centred dipoles, one might wonder if off-centred models
could be tested by measuring the pulsar rate of energy loss.  In
simple cases, the loss of angular velocity $\dot\omega$ of a pulsar
can be described by the law 
\beq
\label{braking}
\dot\omega = - k \omega^n\,,
\eeq
where $n\,$, called the {\it braking index}, 
can be directly inferred  from the instantaneous rate 
of change of frequency through the relation  
$n = {\ddot\omega\omega}/{\dot\omega^2 }$.
A magnetic dipole $\mu$ located on the 
spin axis radiates electromagnetic energy at
the expense of rotational energy at a rate 
\beq
\label{dipole}
\frac{dE_{\mu}}{dt}
=-\halff \frac{d (I \omega^2)}{dt} 
=\frac{2\mu_\rho^2}{3c^3}\>\omega^4\,,   
\eeq
where $I$ is the neutron star moment of inertia.  For constant $I$
this yields $n=3$.  For gravitational quadrupole radiation, denoting
by $M_\rho$ the component of the mass-quadrupole orthogonal to the
spin axis, $dE_M/dt=(G_N M_\rho^2/45c^5)\>\omega^6 $ is found
\citep{ost1969}, and the braking index is $n=5$.  For a magnetic
dipole displaced from the spin axis by an amount $\delta_\rho R\,$,
the spin-down law reads
\beq
\dot\omega=-k \,\omega^3
\left(1+\alpha \omega^2\right)
\eeq
where $k={2\mu^2_\rho}/{3Ic^3}$ and $\alpha = {2\mu^2_z (\delta_\rho
R)^2}/{5\mu^2_\rho c^2}$.  This yields a braking index $ n =
{\ddot\omega\omega}/{\dot\omega^2}\simeq 3+2\,\alpha\omega^2 > 3$.
Unfortunately, for typical values of the relevant parameters ($P\sim
10\,$ms, $R\sim 10\,$km, etc.) one obtains $\alpha\omega^2\lsim
10^{-5}\,$, and thus the deviation from the centred dipole spin-down
law is by far too small to be measurable.

\section{RESONANT SPIN-FLAVOUR PRECESSION}

The mechanism we want to explore to account for pulsar acceleration is
the RSFP of an active neutrino into a sterile neutrino, driven by the
anisotropic magnetic field of a decentred dipole configuration.  For
simplicity, we will not analyze here the more realistic case of an
off-centred magnetic dipole or of more complicated configurations,
since these generalizations do not change our main results.  We just
note that, in the off-centred case, sterile neutrino emission could
contribute not only to the acceleration, but also to a spin-up of the
PNS. As has recently been suggested \citep{spr1998} the rotating core
of the pulsar progenitor might have just a tiny fraction of the
angular momentum required to explain the observed rotation velocities
of pulsars. For this reason it was conjectured that the same physical
process that kicks the PNS at birth could be responsible also for
pulsars' fast rotations \citep{spr1998,cow1998}.  Thus, it might be
interesting to study if a more general magnetic field configuration,
in conjunction with neutrino RSFP, could also account for the observed
pulsars periods.

We start by assuming that $\bar \nu_\tau \rightarrow \nu_s$ RSFP can
occur at neutron star core densities and magnetic field
strengths. Since $\nu_s$ can freely escape from the core, we need to
ensure that the neutrino burst duration $\sim 10\,$s will not be
shortened too much.  Assuming equal luminosity for all neutrino and
antineutrino flavours \citep{jan1995}, even a 100\% efficient
conversion of $\bar \nu_\tau$ into $\nu_s$ would affect only $1/6 \sim
17$\% of the energy in the game. Since there are resonance regions
where the adiabaticity conditions are not satisfied and no conversion
occurs, the overall efficiency for $\bar \nu_\tau \rightarrow \nu_s$
conversion is in fact lower, meaning that the amount of energy carried
away by $\nu_s$ is small enough for the burst duration not to be
drastically changed.  On the other hand a 10\%--20\% asymmetry in the
active--sterile conversion efficiency between opposite hemispheres
would result in a few per cent asymmetry in the total momentum, as is
needed to explain the largest velocities observed.  An important point
in this discussion is to ensure that transitions that involve other
active flavours, such as $\bar\nu_{e,\mu}\to \bar\nu_\tau \to
\nu_s\,$, will not result in an energy `siphon' effect that would cool
the PNS too rapidly.  Indeed, in the light of the present experimental
hints for non-vanishing neutrino mixings, accounting for these
reactions is mandatory.  The rates for neutrino flavour conversion in
a neutron star core were studied by Raffelt and Sigl (1997) and by
Hannestad et al. (1999).  For $\nu_e$ the large charged current (CC)
refractive effects due to the background electrons strongly suppress
any in-matter mixing angle, so that conversion to other neutrino
species cannot occur on the time scale of neutrino diffusion
\citep{raf1993}.  Also for $\nu_\mu \leftrightarrow \nu_\tau$ the
conversion time scale safely exceeds the neutrino diffusion time
\citep{hann1999}. This is due to the presence of a small amount of
muons in the hot superdense core, which implies additional CC
contributions to the $\nu_\mu$ index of refraction, as well as to the
second-order difference in the $\nu_\mu$--$\nu_\tau$ neutral current
(NC) potentials $ \delta V \sim G^2_F m^2_\tau N_N$ (where $N_N$ is
the density of nucleons) \citep{bot1987f}.  In conclusion, the large
differences in the neutrinos indices of refraction inside the PNS
guarantee that $e$, $\mu$ and $\tau$ lepton numbers are separately
conserved, and thus the conversion into $\nu_s$ will affect just one
neutrino flavour.

The condition for a RSFP of $\bar \nu_\tau$ into sterile neutrinos
$\nu_s$ reads
\beq
\label{resonant}
V(\bar\nu_\tau) = 
\frac{G_F N_n}{\sqrt{2}} 
 - \sum_{f=n,p,e}\kappa_f \langle \lambda _\parallel^f \rangle
  =\frac{\Delta m^2}{2 E}\,.
\eeq
The effective potential $V(\bar\nu_\tau)$ felt by a $\bar\nu_\tau$
propagating in the hot, superdense and slightly polarized PNS matter
has two contributions: the first one is due to the coherent vector
interaction with the background and is proportional to the number
density of neutrons $N_n$. The second one, which is due to coherent
axial--vector interactions, is proportional to the average
polarization $\langle \lambda_\parallel^f\rangle $ of the $f=n,p,e$
background fermion parallel to the direction of the neutrino motion
\citep{nun1997,ber1999f}.  The proportionality factor $k_f \sim
(g_A^f\,G_F/\sqrt{2}) N_f$ depends on the$f$-fermion number density
$N_f$ and on its axial--vector coupling $g_A^f$.  In the right-hand
side of equation (\ref{resonant}), $\Delta m^2$ denotes the square
mass difference between the tau and the sterile neutrino masses.  In
the PNS core $V(\bar\nu_\tau)\approx \,4\times 10^{-6}\rho_{14}\,$MeV,
where $\rho_{14} = \rho/10^{14}\,$g cm$^{-3}\,$.  For neutrino thermal
energies of the order $E_\nu \sim 100$--$200\,$MeV the resonance
condition can be satisfied if $\Delta m^2 \sim (20$--$50\,$keV)$^2$.
This value is much larger than the cosmological limit $m_{\nu_\tau}$
($\lsim 10^2\,$eV), and in the following we will therefore assume
$\Delta m^2 = m^2_{\nu_s}-m^2_{\nu_\tau}\simeq m^2_{\nu_s}$.

The average background polarization contributing to the second term in
$V(\bar\nu_\tau)$ grows linearly with the magnetic field, which in a
neutron star can reach extremely high values, up to $\sim 10^{15}\,$G
\citep{tho1992-1993-1995-1996}.  However, during the early cooling
phase when most of the neutrinos are emitted, the temperature is also
large, the electrons are relativistic and degenerate, and as a result
the induced polarizations are strongly suppressed, down to $\langle
\lambda_{n,p} \rangle \lsim 10^{-3}\times B_{15}$ and $\langle
\lambda_e \rangle \lsim 10^{-2}\times B_{15}$ \citep{ber1999}, where
$B_{15}$ is the magnetic field strength in units of $10^{15}\,$G.
Thus in general $\kappa_f \langle \lambda_\parallel^f \rangle \ll {G_F
N_n}/{\sqrt{2}}\,$ and, in most cases, neglecting the polarization
term is well justified.

The resonance condition for the (helicity-conserving) MSW effect
$(\bar\nu_\tau \leftrightarrow \bar\nu_s)$ has the same form that
equation (\ref{resonant}) except for the fact that the right-hand side
is multiplied by $\cos 2\theta_V\,$, where $\theta_V$ is the vacuum
mixing angle. Because the magnetic dipole configuration is asymmetric,
in different regions of a core shell of given neutron density $N_n$
the magnetic field and the average polarization
$\langle\lambda^f\rangle$ can be vastly different. Then it could be
possible that the resonance condition is satisfied with sufficient
accuracy in just one hemisphere.  This can be more easily achieved for
resonant $\bar\nu_e \leftrightarrow\bar\nu_s$ oscillations, since in
this case the neutron number density in the first term in equation
(\ref{resonant}) is replaced by $N_n - 2 N_e\,$, and in some region of
the core and at some stage of the PNS evolution this combination can
approach values close to zero.  However, local violation of the
adiabaticity condition for a RSFP conversion (see below) appears to be
more natural than an asymmetric MSW resonant conversion, in the sense
that it can be satisfied for a larger range of the relevant
parameters, and is also more stable with respect to core evolution
during the deleptonization and cooling phases.  Therefore in the
following we will concentrate on the RSFP conversion mechanism.

\begin{figure}
\epsscale{0.9} \plotone{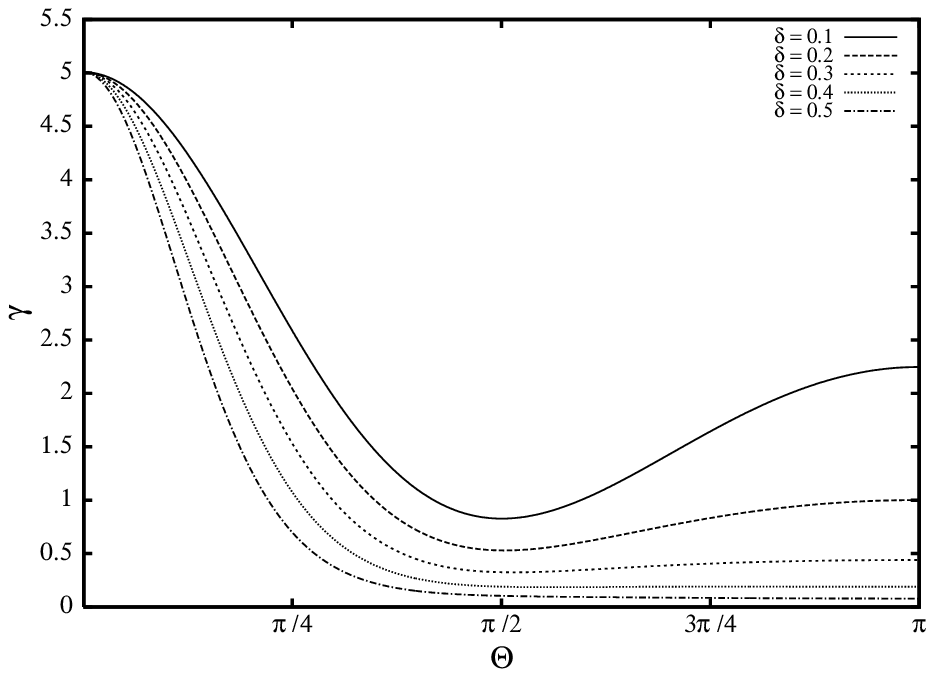} 
\figcaption{\label{gammafig}
Variation of the adiabaticity parameter $\gamma$ inside the resonance
layer as a function of the angular distance $\Theta$ from the magnetic
dipole axis for different values of the decentred parameter
$\delta\,$.  Solid line: $\delta=0.1$; dashed: $\delta=0.2$;
space-dotted: $\delta=0.3$; dotted: $\delta=0.4$; dash-dotted:
$\delta=0.5\,$.  The resonance is located at $\rres=1.5\, R_c\,$ and
the maximum value of the magnetic field strength at the resonance is
$B=4\times 10^{15}\,$G.}
\end{figure}

Once the condition in equation (\ref{resonant}) is satisfied, the
probability $P_{\bar\nu_\tau \leftrightarrow \nu_s}$ that a
spin-flavour neutrino transition will occur depends on its degree of
adiabaticity.  To a good accuracy
\beq
P_{\bar\nu_\tau \leftrightarrow \nu_s} 
\simeq 1- \exp\left(-\frac{\pi}{2}\>\gamma\right)\,,
\eeq
which is sizeable when $\gamma \gsim 1$.  Denoting by $\ell_\rho\equiv
|(1/\rho)(d\rho/dr)|^{-1}$ the characteristic length over which the
density varies significantly, and with $\ell_{\rho\,{\rm res}}$ its
value at the resonance, the adiabaticity parameter $\gamma$ can be
written as \citep{akhm1997,akh1997}
\beqa
\nonumber
\gamma&\simeq& 1.04\left(\frac{1}{1-Y_e}\right)
            \left(\frac{2.6\times 10^{14}\,{\rm  g\,cm}^{-3}}{\rho}\right)
  \\ \label{gamma}
&\times& \left(\frac{\mu_\nu}{10^{-12}\mu_B}\>
       \frac{B_{\perp {\rm res}} }{3\times 10^{15}\,{\rm G}}\right)^2
       \frac{\ell_{\rho\,{\rm res}}}{10\,{\rm km}}\,,  
\eeqa
where $Y_e\approx 0.3 $ is the number of electrons per baryon in the
PNS core \citep{bur1986}, $\mu_B$ is the Bohr magneton and
$B_{\perp}^{\rm res}$ is the value at resonance of the magnetic field
component orthogonal to the direction of neutrino propagation.  Let us
now assume that the magnetic dipole is displaced from the star centre
by an amount $\delta$ in units of the PNS radius $R\,$, and let us
define $\hat\delta=\delta\cdot (R/\rres)\,$, where $\rres$ is the
location of the resonance layer.  For reasonable values $\hat\delta$
$\approx 0.2$--$0.5\,$ the ratio of field intensities between the
north (N) and south (S) magnetic poles of the resonance shell
$B_S^{\rm res}/B_N^{\rm res}\simeq (1-\hat\delta)^3/(1+\hat\delta)^3$
falls in the range $10^{-1}$--$10^{-2}$.  Then from equation
(\ref{gamma}) we see that the condition $\gamma_N \gsim 1$
($P_{\bar\nu_\tau \leftrightarrow \nu_s} \sim 1$) and $\gamma_S \ll 1$
($P_{\bar\nu_\tau \leftrightarrow \nu_s} \sim 0$) that triggers the
anisotropic neutrino emission can be realized in a natural way.  In
Fig. \ref{gammafig} we depict the variation of the adiabaticity
parameter $\gamma$ inside the resonance layer as a function of the
angular distance $\Theta$ from the magnetic dipole axis.  The
resonance is located at $\rres=1.5\, R_c\,$, where $R_c$ denotes the
core radius, and the maximum value of the magnetic field strength at
resonance is $B=4\times 10^{15}\,$G.  For small values of the
decentred parameter $\delta\sim 0.1$ the adiabaticity condition is
matched in both hemispheres and no large asymmetry can be expected.
However, for $\delta > 0.2$ we get $\gamma < 1$ in the whole
hemisphere $\Theta > \pi/2$ and a sizeable asymmetry can be produced.
It is also apparent that if $\delta$ becomes too large, the region
where the adiabaticity condition is satisfied shrinks down to a small
cone (e.g. for $\delta \gsim 0.5$ there is good adiabaticity only when
$\Theta\lsim \pi/6$).  This heavily reduces the conversion efficiency
and results in a suppression of the asymmetry. It also suggests that
no simple correlation between the value of $\delta$ and the overall
$\nu_s$ momentum asymmetry can be expected.

\placefigure{gammafig}

Let us now comment briefly on the possible values of the
relevant physical quantities appearing in equation (\ref{gamma}).  
Magnetic field strengths  $\gsim  10^{15}\,$G are not  
unreasonable  in the interior of a new-born neutron star. 
Indeed, there is observational evidence for highly magnetized 
young pulsars (magnetars) with dipole {\it surface} fields 
as large as $8\times 10^{14}\,$G \citep{hur1999}.
On the other hand,  theoretical studies 
indicate that internal dipole fields as large as  
$B_{\rm int} \sim\> $(5--10)$\>\times 10^{15}\,$G can be  formed 
during the first few seconds after gravitational 
collapse, when the strong convective motions 
coupled with rapid rotation can produce an 
efficient dynamo action \citep{tho1992-1993-1995-1996}. 
As regards the value of $\mu_\nu\,$, 
it is well known that the RSFP conversion mechanism 
requires rather  large transition magnetic 
moments. For densities close to  nuclear density 
$(\approx 2.6\times \rho_{14})$ and acceptable values of 
the magnetic field, the adiabaticity condition in equation (\ref{gamma}) 
points towards magnetic moments of the order of $10^{-12}\mu_B$. 
For $\bar\nu_\tau$ this is several orders of magnitude 
below the laboratory limits \citep{groom1998}.
In the early Universe, $\nu_s$ 
will attain thermal equilibrium only for  
$\mu_\nu\gsim 6\times 10^{-11}\mu_B$ \citep{elm1997}  
and hence the nucleosynthesis constraints can be easily 
satisfied. The astrophysical limit $\mu_\nu\lsim 3\times 10^{-12}\mu_B$
implied by stellar energy-loss arguments 
\citep{raf1990-1999}
does not apply in this case, since the final-state 
sterile neutrino with $m_s\gg 5\,$keV is too heavy to 
be produced inside red--giants or white--dwarfs.
Assuming, as is reasonable, that the main $\nu_s$ 
decay mode is radiative $\nu_s\to \bar\nu_\tau \gamma\,$,
the limits on $\gamma$-ray fluence immediately after the 
arrival of the neutrino burst from the SN1987A  
can be used to set strong constraints on $\mu_\nu\,$.
Under the assumption that the decaying neutrinos are carrying away 
about 1/3 of the total energy, the limit 
$\mu_\nu < 1.6\times 10^{-14}\mu_B \> (10\,{\rm keV}/m_{\nu_s})$ 
was derived \citep{obe1993,raf1996}. However, 
this limit cannot be straightforwardly applied to our case. 
As we will see below, depending on  the overall conversion 
efficiency, the $\nu_s$ can easily carry an overall energy more than 
one order of magnitude smaller than what was assumed in order 
to derive this limit.
It is even conceivable that no conversion at all could have occurred 
for  SN1987A,  if the magnetic field  was too weak to satisfy the 
adiabaticity condition  
(of course this would also imply no intrinsic kick 
for the SN1987A residue, which unfortunately has not been detected). 
In addition to this, in our framework  the $\nu_s$ emission 
is strongly anisotropic, and beaming effects could have 
drastically  reduced the flux along the line of 
sigh of the Earth.
We conclude that the SN1987A limit cannot exclude values 
of $\mu_\nu$ of the order $10^{-12}\mu_B$. 
However, it is interesting to note that a clear signature of 
the mechanism we are proposing would be a large flux of $\gamma$ rays with 
energies of several tens of MeV from the next galactic SN.
Since the $\nu_s$ are emitted from regions close to the hot inner core, 
this would also be accompanied by an anomalous 
spectral component of $\bar\nu_\tau$ from $\nu_s$ decays 
with a temperature about one order of magnitude larger than 
the $\nu_\tau$ and $\nu_\mu$ spectra.

If $\mu_\nu\gsim$ few $\times 10^{-12}\mu_B$
the sterile neutrinos would be directly produced 
through  helicity flipping scattering off electrons and 
protons \citep{bar1989,aya2000f}. 
The main concern here is not so much the 
change in the rate of energy loss,  
since this would affect only one flavour, 
but rather the fact that spin-flips due to scattering 
processes would then occur within the entire core volume with 
an (almost) isotropic distribution.
If the time scale for these processes is much shorter 
than the $\bar\nu_\tau$ diffusion time to the 
resonant region, almost all the $\bar\nu_\tau$ 
will convert into $\nu_s$ inside the core. However, 
in crossing the resonance layer, the $\nu_s$ 
will be anisotropically reconverted into interacting states.
This would result in  approximately the same momentum asymmetry,  
but in the opposite direction.
If the time scale for the helicity flipping scatterings 
is much larger  than the $\bar\nu_\tau$  
diffusion time to the resonance (but of course shorter than 
their diffusion time to the neutrinosphere) the  $\nu_s$ 
asymmetry generated at the resonance surface will be preserved.
This is because the unconverted $\bar\nu_\tau$ 
streaming out from the resonance layer 
will quickly recover an isotropic 
momentum distribution, so that the later emission of   
$\nu_s$ from  helicity flipping scatterings will 
be essentially symmetrical. 
Still, allowing for these possibilities would 
unnecessarily complicate our analysis. Therefore,   
to ensure that helicity flipping scatterings are safely 
suppressed and will not interfere with resonant 
conversions, we will assume that the limit 
$\mu_\nu \lsim$ few $\times 10^{-12}\mu_B\,$ 
\citep{bar1989,aya2000} is satisfied.

\section{THE PROTO-NEUTRON STAR MODEL} 
 

In this section we describe the model that was 
used to simulate the PNS physical conditions 
during the first few seconds after core bounce. 
Our PNS model is three dimensional but static, 
so that the results we will obtain should be 
understood as a ``proof-of-principle''
calculation of the possible size of the 
momentum asymmetry. We identify four different 
regions that are of major importance 
for the conversion process and for the
neutrino emission:

\begin{itemize}

\item[1)] 
{\it Core--atmosphere interface} ($r=R_c$).
Neutrinos are mainly produced within the hot and dense 
core, where the density varies slowly 
with the radius.
From the central regions with supernuclear  
density $\rho_o \simeq 8\times \rho_{14}$ 
the density decreases down to nuclear 
density   $\rho_c\simeq 2.6\times \rho_{14}$.
We define the core--atmosphere interface 
radius $R_c$ through $\rho(R_c)=\rho_c\,$, where   
$R_c\approx 10\,$km.

\item[2)] 
{\it Resonance layer} ($r=\rres$).
This is the region where the (energy-dependent) resonance 
condition in equation (\ref{resonant}) is satisfied.
When the adiabaticity condition $\gamma\gsim 1$ is  
fulfilled, this is also the region 
of emission of the sterile neutrinos. 
In our simulation we have studied the range 
$\rres/R_c = $0.8--1.8,  namely we have 
assumed that the resonance layer is close to the 
core--atmosphere interface.

\item[3)] 
{\it Neutrino energy--sphere} ($r=R_E$). 
The main reactions through which the 
$\bar\nu_\tau$ exchange energy with 
the stellar gas are 
NC scattering off electrons and neutrino 
pair processes. The neutrino energy sphere 
is defined as  the region where the 
$\bar\nu_\tau$ undergo the last inelastic 
interaction with the background particles. 
For $r>R_E\,$, neutrinos mainly scatter off nucleons. 
While this determines the neutrino transport 
opacity, the amount of energy exchanged is negligible,  
so that outside the energy--sphere  
the neutrinos start being thermally 
disconnected from the medium.

\item[4)] 
{\it Neutrino transport--sphere (neutrinosphere)} ($r=R_T$). 
This is the layer with optical depth~$\approx 1$ 
which is determined by the neutrino elastic
cross section off nucleons. The cross 
section depends on the square of the neutrino energy,  
and thus $R_T$ (as well as $R_E$) is an energy-dependent quantity.  
Since $R_T\gg \rres$ the initial asymmetry generated at $\rres$ 
vanishes at $R_T$ because of multiple scatterings that  
completely redistribute the momentum direction of the 
unconverted $\bar\nu_\tau$. 
For $r>R_T$ the neutrinos no longer undergo 
diffusion processes and freely escape from the star.  

\end{itemize}

\begin{figure}
\epsscale{0.9}
 \plotone{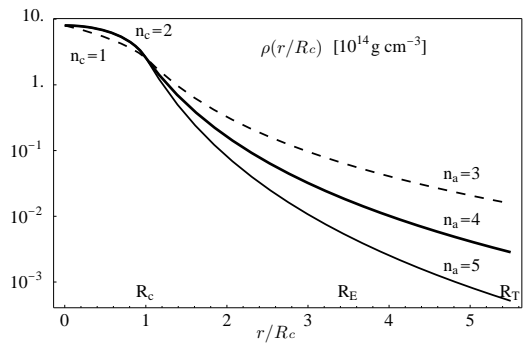}
\figcaption{\label{densityfig}
Density profiles in units of $10^{14}\,$g cm$^{-3}$
as a function of  $r/R_c\,$, for different 
values of the exponents $n_c$ and $n_a$ 
[see Eq. (\ref{density})]. 
The dark solid line depicts the profile that was used in the 
simulations, and  corresponds to $n_c=2$ and $n_a=4\,$, 
the dashed line corresponds to $n_c=1$ and $n_a=3\,$, and
the light solid line corresponds to $n_c=2$ and $n_a=5\,$.
The position of the core--atmosphere interface $(R_c)$ and 
the approximative positions of the $\bar\nu_\tau$ energy--sphere 
($R_E$) and transport--sphere ($R_T$) are also shown.}
\end{figure}

The density and temperature profile
for the PNS  core and atmosphere 
that have been used in our simulations are given 
by simple analytical formulae.
These expressions   
have been chosen according to the following criteria:
(i) 
they approximately reproduce 
the profiles resulting from detailed  
numerical studies of the PNS 
evolution \citep{bur1986};
(ii)
they satisfy the physical requirements   
that the $\bar\nu_\tau$ decouple energetically 
with a spectral temperature $T(R_E)\approx 8\,$MeV and  
around densities $\rho(R_E) \sim 10^{12}$ g cm$^{-3}$; 
(iii) 
they result in conservative estimates  
of the overall asymmetry in the sterile 
neutrino emission.  The density profile is  
given by 
\beq
\label{density}
\rho(r) = \cases{
\rho_0 \left[ 1 - \left( 1 - \frac{\rho_c}{\rho_0}\right) 
\left( \frac{r}{R_c}\right)^{n_{c} }\right]  & if  $r \leq R_c\,$; \cr
\rho_c \left( \frac{R_c}{r}\right)^{n_{a} } & if  $r > R_c\,$.}
\eeq
For suitable values of the exponents $n_{c}$ for the core 
and $n_{a}$ for the atmosphere, this expression 
reproduces reasonably well  
the results of a detailed numerical computation  
\citep{bur1986}, namely a slowly varying 
density for the core ($n_c\sim 1,2$) and a steeper 
profile $\sim r^{-n_{a}}$ (with $n_{a}\sim 3$--5)  
for the atmosphere. 
In Fig. \ref{densityfig} we depict a set of profiles 
for different values of $n_{c}$ and $n_{a}$. 
In our simulation we have used  
$n_{c}=2\,$, which gives an appreciable 
density variation inside the core. 
With respect to a milder variation 
or to a constant profile, this is a 
conservative choice, since it favors the  outwards 
diffusion of $\bar\nu_\tau$ and 
reduces the probability of random  
crossings of the adiabatic resonance region. 
For the atmosphere we used $n_{a}=4\,$, which enhances   
with respect to steeper profiles, the  
probability  that some of the neutrinos will diffuse 
back from the atmosphere and cross the resonance region 
in  the  `wrong'  direction.

\placefigure{densityfig}

For the temperature profile we assumed a simple 
model with an  isothermal core  
\beq
\label{temperature}
T = \cases{ 
T_o  & if $r\leq R_c\,$; \cr
T_o \left( \frac{R_c}{r}\right)^{n_T} & if $r > R_c\,$, }
 \eeq
with $T_o = 30\,$MeV\citep{bur1986} and $n_T=1$.
Since the $\nu_s$ are emitted with spectral temperatures
close to the core temperature, larger values of $T_o$ would 
enhance the momentum asymmetry.   
During the early stage of the PNS formation, 
the rapid deleptonization process  produces changes 
in the temperature on the time scale of hundreds 
of milliseconds. In the first few seconds    
after core bounce, neutrinos are emitted predominantly 
from the outer regions where the density is lower.
Matter loses entropy, undergoes compression,  
and the temperature increases to values much larger 
than the central temperature. 
The temperature inversion drives the heat flux 
towards the interior;  however, the peak temperature 
reaches the central regions about 10 s
after core bounce, when most of the neutrinos
already escaped from the core \citep{bur1986}.
Since the neutrino opacity grows with the
square of the energy, a temperature inversion 
tends to keep the neutrinos trapped  in the inner 
regions, enhancing the probability for their  
conversion. We have verified that 
profiles with temperature inversion generally  
result in larger momentum asymmetries, and thus the 
isothermal core  represents a  conservative approximation. 
As regards the temperature profile for the atmosphere, 
we assumed that it decreases with the 
radius as $1/r\,$.  With respect to steeper temperature 
gradients this again favors the probability 
of  `wrong' back-crossings of the resonance.


The $\bar\nu_\tau$ elastic scattering 
off nucleons reads 
\beq
\label{elastic}
\frac{d \sigma_E}{d \cos\theta}=
\frac{G^2_F E^2_\nu}{8\pi}
\left(C_1 + C_2\cos\theta\right)\,, 
\eeq
where $\theta$ is the scattering angle between the incoming 
and outgoing neutrino directions, and the coefficients 
$C_1$ and $C_2$ depend on the neutrino--nucleon couplings 
averaged over the $p$ and $n$ number densities.   
For the PNS nuclear matter the canonical values $C_A^p=-C_A^n=1.26$ 
given  by isospin invariance are modified due to strange-quarks contribution 
to the nucleon spin. We use the values 
$C_V^n=-1\,$, 
$ C_V^p=1-4\sin^2\Theta_W\simeq 0.07 \,$,
$ C_A^n\approx-1.15 \,$,
$ C_A^p\approx 1.37$ \citep{raf1995,kei1995f}.
Using also $Y_p=1-Y_n\approx 0.3\,$ for the relative 
abundances of the nucleons, we obtain   
\beqa
\label{averages}
C_1 \!\!\!\!&=&
\!\!\!\!\!\!\! 
     \sum_{N=n,p}Y_N  
\left[\left({C_V^N}\right)^2 + 
3 \left({C_A^N}\right)^2\right] \approx 5.2\,, \\
C_2 \!\!\!\! &=& \!\!\!\!\!\!\! 
     \sum_{N=n,p}Y_N 
\left[\left({C_V^N}\right)^2-
\left({C_A^N}\right)^2\right] \approx -0.8.  
\eeqa
Equation (\ref{elastic}) determines the $\bar\nu_\tau$ transport
opacity. Inelastic reactions where the energy exchange is of order
unity, such as neutrino--electron scattering and neutrino-pair
processes, give only a minor contribution to the total opacity.
However, they are responsible for keeping the neutrinos in thermal
equilibrium with the background, and determine the relative positions
of $R_T$ and $R_E$.  We have assumed a relative rate between inelastic
and elastic processes of the order of 10\%, so once every ten
scatterings we generate again the neutrino energy according to the
neutrino position within the star and to the corresponding local
temperature.  Since the $\bar\nu_\tau$ emerge from the resonant layer
with an overall momentum asymmetry equal and opposite to that of the
$\nu_s\,$, their thermalization and especially the redistribution of
their momentum due to multiple scattering is crucial for ensuring that
the $\nu_s$ momentum asymmetry is left unbalanced.

\section{RESULTS}

Neutrinos are generated randomly inside the core with a Fermi-Dirac
distribution corresponding to a spectral temperature $T_o=30\,$MeV and
zero chemical potential, and are left diffusing outwards. Of course,
the loss of $\bar\nu_\tau$ due to conversion into sterile neutrinos
eventually builds up a non-vanishing chemical potential.  However,
since our simulation of the neutrino diffusion is static (that is
there is no time evolution of the relevant parameters characterizing
the PNS model) this effect has been neglected.  
To study the effect of the decentred magnetic dipole, we have run
simulations with $\delta$ ranging between $0.1$ and $0.6$.  We have
also studied the effect of moving the resonance layer from inside the
core ($\rres/R_c=0.8\,$, $\rho_{\rm res}\approx 4.5\times \rho_{14}$,
$m_{\nu_s}\approx 50\,$keV) to three different locations in the
atmosphere $\rres/R_c=1.3\,,1.5\,$ and $1.8$ (the last value
corresponds to $\rho_{\rm res}\approx 0.25\times \rho_{14}$,
$m_{\nu_s}\approx 12\,$keV).  The reference value for the neutrino
magnetic moment has been fixed at $\mu_\nu = 10^{-12}\mu_B$ (of course
rescaling $\mu_\nu \to k\mu_\nu$ and $B \to B/k$ would leave the
results unchanged) and we have varied the magnetic field in the star
interior between $3\times 10^{15}\,$G and $8\times 10^{15}\,$G.  The
values of $B$ correspond to maximum values of the adiabaticity
parameter at the resonance surface in the range $\gamma\approx 3$--12.
In contrast, owing to the anisotropies of the magnetic field, in the
resonance regions far from the dipole centre, $\gamma\ll 1$.  Once the
$\bar\nu_\tau$ reach the resonance, the component of the magnetic
field transverse to the direction of propagation is computed, and they
are converted into $\nu_s$ with the appropriate transition probability
$P_{\bar\nu_{\tau}\leftrightarrow \nu_s}$.  The $\nu_s$ escape from
the star without interacting further.  The large differences in the
spatial values of $\gamma$ can result in huge anisotropies in the
conversion efficiency between different points of the resonance layer,
so that the $\nu_s$ are preferentially emitted from one hemisphere.
In computing the $\nu_s$ momentum asymmetry we have neglected the
effect of gravitational redshift.  On the opposite hemisphere, most of
the $\bar\nu_\tau$ are not converted; however, further interactions
with the medium wash out their asymmetry almost completely.  Of
course, from regions close to the core--atmosphere interface, some
$\bar\nu_\tau$ can scatter inwards and resonantly convert when
crossing the resonance region, so that a certain number of $\nu_s$
will be emitted in the `wrong' direction, thus reducing the overall
asymmetry.  When the resonance lies inside the core, this effect is
somewhat enhanced since a sizeable fraction of $\bar\nu_\tau$ are
produced in the region $\rres<r<R_c\,$.

The fractional asymmetry in the 
total momentum of the neutrino emission 
needed to accelerate the remnant 
at a velocity  $V$  is given by 
\beq
\label{kick}
\frac{\Delta p}{p_{\rm tot}}\simeq
0.03\, \left(\frac{V}{10^3\,{\rm km\> s}^{-1}}\right)\,
      \left(\frac{E_{\rm tot}}{3\times 10^{53}\>{\rm ergs}}\right)\, 
\left(\frac{M}{1.5\,M_\odot}\right)\,, 
\eeq
where $E_{\rm tot}$ is the total energy 
radiated in neutrinos and $M$ is the PNS mass. 
Assuming equipartition of the total luminosity 
between the active flavours, we see that 
to explain velocities of several hundreds of 
km s$^{-1}$ the total asymmetry in the momentum 
carried by the $\nu_s$ and by the surviving 
$\bar\nu_\tau$ $(p(\nu_s+\bar\nu_\tau)\approx 0.17\, p_{\rm tot})$  
should amount to  $\approx 10$\%--20\%. 
The results of different simulations 
run with a statistics of $\approx 4\times 10^5$ $\bar\nu_\tau$  
are presented in Table~I. 
It is apparent that momentum asymmetries of the correct size 
can be generated. From the table it 
is also possible to infer the impact 
that the different parameters have on the size of the asymmetry.
As expected, increasing the value of the  decentred parameter 
decreases the fractional number of $\nu_s$ produced. 
The overall asymmetry first increases; however,  
when the conversion efficiency decreases too much, 
also the asymmetry gets reduced. 
If the resonance is inside the core, the asymmetries 
are generally small. This is mainly  
due to the large number of `wrong' crossing of the
resonance region,  and to some extent also to the moderate 
value of $\gamma_{\rm max}$ (larger values would require 
raising the magnetic field up to $B\approx 10^{16}\,$G). 
We stress that a crucial condition that 
has to be satisfied to get large asymmetries is that 
the adiabaticity parameter $\gamma$ should reach  
values  sizeably larger than unity.
In fact if $\gamma_{\rm max} \sim 1$ the mechanism is rather 
inefficient, since only the $\bar\nu_\tau$ with  momentum almost 
orthogonal to the magnetic field direction will convert. 
Besides yielding a small conversion efficiency, 
this also implies that the momenta of the 
emerging $\nu_s$ will be mainly oriented tangentially to the 
resonant layer and  will not contribute to build up an    
asymmetry. Only for $\gamma_{\rm max} \gsim 3$--4,  
a sizeable momentum component in the radial direction 
can be generated.

\begin{deluxetable}{ccccccc}  
\tablecolumns{7}
 \tabletypesize{\footnotesize}  
\tablewidth{380pt}  
\tablecaption{Results\label{table}}  
\tablehead{  
  \colhead{$\frac{\rres}{R_c}$} & \colhead{$ B$} & \colhead{$ \delta=\frac{R_B}{R_c} $}                  &
  \colhead{$\frac{N({\nu_s})}{N({\bar\nu_\tau+\nu_s})} $}   & \colhead{$\frac{\Delta p}{p}\,(\nu_s)$}    & 
  \colhead{$\frac{\Delta p}{p}\,(\bar\nu_\tau)$} & \colhead{$\frac{\Delta p}{p} (\bar\nu_\tau + \nu_s)$} } 
\startdata  
0.8        & $8 \times10^{15}\,{\rm G} $      & 0.1 & 0.58 &0.03  &--0.001  & 0.02  \\
\          & $(\gamma_{\rm max} \simeq 3.3)$  & 0.2 & 0.43 &0.06  &--0.002  & 0.05  \\
\          &                                  & 0.3 & 0.27 &0.09  &--0.004  & 0.04  \\
\          &                                  & 0.5 & 0.08 &0.14  &--0.004  & 0.03  \\ \tableline
1.3        & $6 \times10^{15}\,{\rm G}$       & 0.2 & 0.57 &0.11  &--0.002  & 0.09  \\
\          & $(\gamma_{\rm max}\simeq 5.5 )$  & 0.3 & 0.45 &0.21  &--0.010  & 0.15  \\
\          &                                  & 0.4 & 0.33 &0.29  &--0.006  & 0.17  \\
\          &                                  & 0.6 & 0.17 &0.44  &--0.004  & 0.17  \\ \tableline
1.5        & $6 \times10^{15}\,{\rm G}$       & 0.3 & 0.51 &0.13  & +0.003  & 0.10  \\
\          & $(\gamma_{\rm max}\simeq 11.3)$  & 0.4 & 0.48 &0.21  &--0.005  & 0.16  \\
\          &                                  & 0.5 & 0.44 &0.27  &--0.002  & 0.16  \\
\          &                                  & 0.6 & 0.29 &0.37  &--0.002  & 0.21  \\ \tableline
1.5        & $4 \times10^{15}\,{\rm G}$       & 0.2 & 0.54 &0.11  &--0.005  & 0.09  \\
\          & $(\gamma_{\rm max}\simeq 5.0)$   & 0.3 & 0.44 &0.19  & +0.001  & 0.12  \\
\          &                                  & 0.4 & 0.37 &0.26  &--0.006  & 0.17  \\
\          &                                  & 0.5 & 0.26 &0.36  &--0.003  & 0.15  \\ \tableline
1.8        & $3 \times10^{15}\,{\rm G}$       & 0.2 & 0.60 &0.09  &--0.011  & 0.07  \\
\          & $(\gamma_{\rm max}\simeq 7.0)$   & 0.3 & 0.58 &0.12  & +0.006  & 0.09  \\
\          &                                  & 0.4 & 0.43 &0.22  & +0.003  & 0.15  \\
\          &                                  & 0.6 & 0.28 &0.39  & +0.006  & 0.10  \\
\enddata  
\tablecomments{The results of our simulations. Different positions of the resonance 
layer have been studied: inside the core ($\rres/R_c=0.8$) and outside the core  
($\rres/R_c=1.3\,,1.5\,,1.8$). The values of the magnetic field are given in the 
second column, together with the corresponding maximum value of the adiabaticity 
parameter $\gamma$. In all the simulations we have assumed $\mu_\nu = 10^{-12} \mu_B\,$. 
The displacement of the magnetic dipole from the star centre  $\delta=R_B/R_c$ is listed 
in the third column, while the resulting overall efficiency for $\bar\nu_\tau$ conversion 
into $\nu_s$ is given in the fourth column.  The momentum asymmetry for the $\nu_s$, for 
the unconverted $\bar\nu_\tau,$ and the overall ${\bar\nu_\tau + \nu_s}$ momentum asymmetry 
are given in the last three columns respectively.}
\end{deluxetable}  

If the resonance is just outside the region where most 
of the neutrinos are produced ($\rres\gsim R_c$), 
asymmetries of the correct size are easily obtained.
Because of the lower value of the local density, this  
allows for larger values of $\gamma_{\rm max}$ and   
also requires somewhat smaller magnetic fields. 
However, if  $\mu_\nu \lsim 10^{-12}\mu_B$ 
internal magnetic field strength $B\gsim 10^{15}\,$G are 
needed in any case.  Since the relevant combination  
that determines the value of $\gamma$ 
is the product $\mu_\nu B$, for larger 
transition magnetic moments the required 
magnetic fields could be smaller by a factor of a few.

\placetable{table}

In spite of this, it is clear from our results  
that no obvious correlation between the
kick velocity and the magnetic field strength 
can be expected. Similarly, any other correlation pattern 
would be quite difficult to recognize.
In fact, for a given value of the neutrino 
transition magnetic moment $\mu_\nu\,$   
the efficiency of the conversion is determined 
by a complicated interplay between the 
position of the resonance layer $\rres$, the value
of the decentred parameter $\delta$, and the 
strength of the magnetic field $B$. 
In some cases a large magnetic field together 
with a large value of $\delta$ implies that   
$\gamma \gg 1$ only in a relatively small 
region, thus lowering too much the conversion efficiency
and implying that the $\nu_s$ emission is mainly 
oriented tangentially to the resonance layer surface.  
Larger conversion rates are  produced by 
smaller values of $\delta$, and sometimes this can 
also result in larger asymmetries.
The exact value of the core temperature can also 
affect the results. For larger temperatures,   
about the same numbers of $\nu_s$ will be produced,   
but with a higher spectral temperature.
This effect tends to increase the asymmetry.

Finally, we should mention that since the $\nu_s$ emission 
is likely to occur on a time scale much larger than the 
pulsar rotational period at birth, the resulting 
momentum kick would be averaged out proportionally 
to the cosine  of the angle between the 
spin axis and the magnetic axis.  
However, even if we assume that the present  
orientation of the magnetic axis is representative 
of its orientation at birth, we believe that  it is not possible to 
predict any  unambiguous correlation between the pulsars'  
velocities and their magnetic axis orientation, because 
of the large number of different parameters that concur to 
determine the overall effect.

In conclusion, we have shown that $\bar\nu_\tau \leftrightarrow \nu_s $
RSFP inside the PNS core is able to account for pulsar  
natal kicks of the required size. 
The crucial assumptions to achieve this result 
are (i)~the presence of sizeable anisotropies in the PNS 
magnetic field configuration; (ii)~internal magnetic field
strengths of the order of few $\times 10^{15}\,$G; 
(iii)~a sterile neutrino with a mass of a few tens  
of keV and with a transition magnetic moment 
with an active  flavour of the order of $10^{-12}\mu_B$. 
The first two assumptions have already been discussed  
in this paper, so let us  comment briefly on the last point. 
If all the active neutrinos have a mass satisfying the 
cosmological limit $\lsim 100\,$eV, it would be somewhat 
complicated to construct a  particle physics model that   
realizes the third condition, since a large magnetic moment  
generally implies rather large radiative contributions to the 
neutrino masses. This problem was extensively addressed 
in the past, and some clever solutions were proposed 
\citep{vol1988,bar1989}. We believe that a consistent  
particle physics model yielding a large $\bar\nu_\tau$--$\nu_s$
transition magnetic moment and a light $\nu_\tau$ can  
be constructed along the same lines. 
Let us also mention that quite recently the cosmological limits 
on neutrino masses have been revisited under the hypothesis 
that the Universe underwent a non-trivial thermal 
evolution right before the nucleosynthesis era \citep{gui2000f}. 
The result of this study opens up the possibility 
that also $m_{\nu_\tau}$ could  be of the order of 
a few tens of keV. Indeed this would render the whole picture
much more natural from the particle physics point of view.

\acknowledgements
This work was supported in part by BID and Colciencias in Colombia
under contract 401-97 code 1115-05-087-97.
We acknowledge E. Akhmedov for useful conversations and 
M. Kachelriess for bringing to our attention the work of 
Oberauer et al. (1993).


\end{document}